%% file: main.tex
\documentclass[sigconf,anonymous=False,review=False,natbib=True]{acmart}
\usepackage{multirow}
\usepackage{tabularx}
\usepackage{graphicx}
\usepackage{xcolor}
\usepackage{subfigure}
\usepackage{array}
\usepackage{balance}
\usepackage{enumitem}
\usepackage{caption}
\usepackage{booktabs}
\usepackage{amsfonts}
\usepackage{geometry}
\input{definitions}

\makeatletter
\def\hlinew#1{%
  \noalign{\ifnum0=`}\fi\hrule \@height #1 \futurelet
   \reserved@a\@xhline}
\AtBeginDocument{%
  \providecommand\BibTeX{{%
    \normalfont B\kern-0.5em{\scshape i\kern-0.25em b}\kern-0.8em\TeX}}}

 \setcopyright{acmcopyright}

\copyrightyear{2022} 
\acmYear{2022} 
\setcopyright{acmlicensed}
\acmConference[WWW '22]{Proceedings of the ACM Web Conference 2022}{April 25--29, 2022}{Virtual Event, Lyon, France}
\acmBooktitle{Proceedings of the ACM Web Conference 2022 (WWW '22), April 25--29, 2022, Virtual Event, Lyon, France} 
\acmPrice{15.00} 
\acmDOI{10.1145/3485447.3511966} 
\acmISBN{978-1-4503-9096-5/22/04}

\begin{document}
\fancyhead{}
\title{Towards a Better Understanding of Human Reading Comprehension with Brain Signals}


\author{Ziyi Ye$^1$, Xiaohui Xie$^1$, Yiqun Liu$^1*$, Zhihong Wang$^1$, Xuesong Chen$^1$, Min Zhang$^1$, \\ and Shaoping Ma$^1$}
\def \authors{Ziyi Ye, Xiaohui Xie, Yiqun Liu, Zhihong Wang, Xuesong Chen, Min Zhang, and Shaoping Ma}
\affiliation{
  \institution{$^1$Department of Computer Science and Technology, Institute for Artificial Intelligence, Beijing National Research Center for Information Science and Technology, Tsinghua University
  \city{Beijing}
  \country{China}}
}
\email{yeziyi1998@gmail.com,xiexh_thu@163.com,yiqunliu@tsinghua.edu.cn,wangzhh629@mail.tsinghua.edu.cn,}
\email{chenxuesong1128@163.com,z-m@tsinghua.edu.cn,msp@tsinghua.edu.cn}
\thanks{$*$Yiqun Liu is the corresponding author.}

\renewcommand{\shortauthors}{Ye, et al.}

\begin{abstract}

Reading comprehension is a complex cognitive process involving many human brain activities.
However, little is known about what happens in human brain during reading comprehension and how these cognitive activities can affect information retrieval process.
Additionally, with the advances in brain imaging techniques such as \ac{EEG}, it is possible to collect brain signals in almost real time and explore whether it can be utilized as  feedback to facilitate information acquisition performance.

In this paper, we carefully design a lab-based user study to investigate brain activities during reading comprehension. 
Our findings show that neural responses vary with different types of reading contents, i.e., contents that can satisfy users' information needs and contents that cannot.
We suggest that various cognitive activities, e.g., cognitive loading, semantic-thematic understanding, and inferential processing, underpin these neural responses at the micro-time scale during reading comprehension.
From these findings, we illustrate several insights for information retrieval tasks, such as ranking models construction and interface design. 
Besides, with the emerging of portable \ac{EEG}-based applications, we suggest the possibility of detecting reading comprehension status for a proactive real-world system.
To this end, we propose a \textbf{U}nified framework for \textbf{E}EG-based \textbf{R}eading \textbf{C}omprehension \textbf{M}odeling~(UERCM).
To verify its effectiveness, we conduct extensive experiments based on EEG features for two reading comprehension tasks: answer sentence classification and answer extraction. 
Results show that it is feasible to improve the performance of two tasks with brain signals. 
These findings imply that brain signals are valuable feedback for enhancing human-computer interactions during reading comprehension.

\end{abstract}



\maketitle
\input{1-introduction.tex}

\input{2-related_work.tex}

\input{3-user_study.tex}

\input{4-statistical_analysis.tex}

\input{5-experiments_and_discussions.tex}

\input{6-conclusion.tex}

\input{7-acknowledgements.tex}

\bibliographystyle{ACM-Reference-Format}
\balance
\newpage
\bibliography{references.bib}
\appendix
\newpage
\input{7-appendix.tex}

\end{document}

%% file: definitions.tex
\usepackage{acronym}

\acrodef{IR}{Information Retrieval}
\acrodef{SERP}{search engine result page}
\acrodef{EEG}{electroencephalogram}
\acrodef{ERP}{event related potential}
\acrodef{MAP}{mean average precision}
\acrodef{nDCG}{normalized discounted cumulative gain}
\acrodef{ERPF}{Event-related-potential-based feature}
\acrodef{MKL}{Multiple Kernel Learning}
\acrodef{SVM}{support vector machines}
\acrodef{RBF}{radial basis function}
\acrodef{GBDT}{gradient boosting regression tree}
\acrodef{LR}{logistic regression}
\acrodef{MLP}{multilayer perceptron}
\acrodef{RNN}{recurrent neural network}
\acrodef{fMRI}{functional magnetic resonance imaging}
\acrodef{SHAP}{SHapley Additive exPlanations}
\acrodef{ROI}{Region of interest}
\acrodef{IN}{Information Need}
\acrodef{MEG}{Magnetoencephalography}
\acrodef{AUC}{Area Under Curve}
\acrodef{IDF}{inverse document frequency}
\acrodef{FBF}{Frequency-band-based feature}
\acrodef{BCI}{brain–computer interface}
\acrodef{CVOT}{10-fold cross validation on tasks}
\acrodef{LOPO}{leave-one-participant-out}
\acrodef{AUC}{Area under the ROC curve}
\acrodef{BL}{baseline}
\acrodef{CRF}{linear-chain Conditional Random Field}
\acrodef{MRR}{Mean Reciprocal Rank}
\acrodef{NLP}{natural language processing}

\AtBeginDocument{%
  \providecommand\BibTeX{{%
    \normalfont B\kern-0.5em{\scshape i\kern-0.25em b}\kern-0.8em\TeX}}}

%% file: 1-introduction.tex
\section{introduction}

Human reading comprehension is a complex cognitive process involved in many search processes, e.g., information seeking and relevance judgment. 
In that regard, understanding reading comprehension is beneficial for a more proactive \ac{IR} system, such as inferring search intent~\cite{gwizdka2014characterizing}, designing search interface~\cite{qu2019answer}, and constructing ranking models~\cite{li2019teach}. 
Prior studies have investigated the behavioral patterns and attention allocation mechanisms of human reading process utilizing mouse movement~\cite{liu2015different} and eye-tracking~\cite{li2018understanding}. 
However, these methods can't straightforwardly uncover the actual cognitive activities in the brain and underlying psychological factors during reading comprehension.
Hence, the question of ``What is the nature of reading comprehension in \ac{IR} scenarios?'' remains an open problem.  

Recently, the rapid developments of neuroimaging technology~(e.g., \ac{EEG} and \ac{fMRI}) make it feasible to explore brain activities in \ac{IR} scenarios. 
Extensive studies have applied neurological devices to explore the emergence of \ac{IN}~\cite{moshfeghi2016understanding} and relevance judgment procedure~\cite{allegretti2015relevance}.
These studies constitute an important step in unraveling some cognitive processes in \ac{IR} scenarios and provide findings that can not be obtained by previous techniques like eye-tracking. 
Nonetheless, few works have thoroughly investigated the cognitive processes during reading comprehension, i.e., neural responses when users locate \textit{key information} for their \ac{IN}.
In this paper, key information refers to answers and semantic-related spans~(examples are shown in Table~\ref{tab:dataset_sample}). 
We believe that understanding these cognitive processes is beneficial for information retrieval tasks, such as ranking models construction and interface design.

On the other hand, recent studies apply brain signals as user feedback for predicting realization of \ac{IN}~\cite{moshfeghi2019towards} and relevance~\cite{gwizdka2017temporal,koelstra2009eeg}. 
With the advances of portable \ac{BCI} equipment, \citet{liu2021challenges} suggest applying \ac{BCI} in real-life settings.
\ac{BCI} can resolve problems in many situations where conventional signals are noisy, e.g., queries are too short or ambiguous, users switch their intents during the search process. 
However, brain signals have rarely been applied to detect users' reading and answer-seeking states, i.e., whether users have found useful part~(answer sentence classification) and whether users have located the answer phrases~(answer extraction). 
These would be helpful for the proactive \ac{IR} systems in the near future.

In this paper, we aim to interpret the cognitive processes during reading comprehension and explore the effectiveness of brain signals to complete reading comprehension tasks, raising the following research questions:
\begin{itemize}
	\item \textbf{RQ1}: Are there any detectable differences in brain activities between reactions to key information and ordinary information during reading comprehension? If yes,
	\item \textbf{RQ2}: What are the cognitive bases of these differences and their insights for IR? And,  
	\item \textbf{RQ3}: Is it possible to classify answer sentences and locate potential answer words with these differences?
\end{itemize}
 
To shed light on these research questions, we conduct a lab-based user study to investigate reading comprehension in the context of question answering.  
In this user study, an \ac{EEG} device is applied to collect brain activities, which are later examined with \ac{ERP} analysis, a typical method in neuroscience~\cite{luck2000event}. 
Based on the analysis, we find brain activities vary with different types of contents. 
Notably, we find that the particular \ac{ERP} component N100-P200, which is associated with cognitive loading~\cite{raney1993monitoring}, differs in answer words, semantic-related words, and ordinary words~(The definition of \ac{ERP} components, e.g., N100-P200 and P400, can be found in Section~\ref{user study:Introduction to ERP}). 
Answer words contribute to larger P600, which is caused by cognitive activities of inferential processing. 
Based on that neural basis, we illustrate several insights for \ac{IR} community.
For instance, (1) the ranking models construction should consider fine-grained document structure to reduce cognitive load and avoid misunderstanding, (2) the result snippet should provide not only semantic-related content but also contextual information for a better understanding.

Furthermore, inspired by the development of \ac{BCI}, we explore the possibility of using brain signals for detecting reading status.
We propose a \textbf{U}nified framework for \textbf{E}EG-based \textbf{R}eading \textbf{C}omprehension \textbf{M}odeling~(UERCM), which can utilize brain signals to complete two reading comprehension tasks: answer sentence classification and answer extraction.
Experimental results show that UERCM leads a significant improvement of 0.179 in answer sentence classification~(in terms of \ac{MAP}) and 0.157 in answer extraction~(in terms of \ac{AUC}) compared to the untrained model, respectively.
It also outperforms other baselines, especially in the answer sentence classification task.

%% file: 2-related_work.tex
\section{related work}
\subsection{Reading Comprehension}
Reading comprehension is a cognitive process for acquiring information in text-based search scenarios, which involves vision processing, semantic understanding, and information gaining~\cite{crowder1992psychology}.
Many prior works study users' reading patterns and attention allocation in \ac{IR} scenarios with eye-tracking devices.
\citet{gwizdka2014characterizing} investigates the reading behavior with eye movements and indicates text document processing depends on relevance and perceived relevance. 
\citet{li2018understanding} study the attention distribution during passage-level reading comprehension with eye movements and explicit feedback.
Then a two-stage reading model is further processed with their findings. 

Moreover, existing works investigate implicit feedback during reading comprehension.
For example, \citet{liu2015different} utilize mouse movements to study the \ac{SERP} inspecting process and predict users' satisfaction with Web pages.
\citet{cole2013inferring} show that eye movement patterns during reading comprehension could infer users' pre-knowledge for better modeling search context.
\citet{zheng2019human} extract eye movements' features derived from reading tasks to improve performance in the machine reading comprehension task.

Although neuroscience technology has been used to study reading behavior in the general domain, e.g., word recognition~\cite{hsiao2011visual} and syntactic analysis~\cite{newman2010effect}, little research literature concentrate on studying reading comprehension in \ac{IR} scenarios, which involve the information seeking process.
Hence, we try to uncover the psychological factors when people perceive key information and understand human reading comprehension from a neuroscience perspective.

\subsection{Neuroscience \& IR}
There is a growing number of researches using neuroimaging techniques to study \ac{IR}-related tasks. 
These works mainly focus on studying the fundamental concepts in \ac{IR}~(e.g., \ac{IN} and relevance) and leveraging brain signals as implicit feedback.   
In terms of \ac{IN}, \citet{moshfeghi2016understanding} use \ac{fMRI} to examine the neural processes involved in how \ac{IN} emerges. 
They reveal a distributed network of brain regions commonly associated with activities related to the \ac{IN}.
Besides \ac{IN}, previous works use brain signals to understand relevance from a neuroscience perspective. 
In particular, \ac{fMRI} devices with a higher spatial resolution are adopted to identify which brain regions are activated~\cite{moshfeghi2013understanding}, while EEG devices with a higher time resolution to determine when relevance judgment is happening~\cite{allegretti2015relevance}.

As implicit feedback, brain signals are widely used in \ac{IR} scenarios, including predicting realization of \acp{IN} and relevance.
\citet{moshfeghi2019towards} propose generalized and personalized methods to predict the realization of \acp{IN} using \ac{fMRI} features. 
Moreover, \citet{kauppi2015towards} conduct a feasibility study on predicting the relevance of visual objects with \ac{MEG}-based classifiers.
For textual information relevance, ~\citet{gwizdka2017temporal} apply eye movements and EEG signals to the assessment of document relevance.
Their classification model with \ac{EEG} features shows an improvement of 20\% in \ac{AUC} compared with random baseline.

Recently, along with the advances of portable \ac{BCI} equipment and motivated from their wide applications in education~\cite{nascimento2021education} and game playing~\cite{parbez2020blinkfruity}, it appears reasonable to utilize brain signals for search performance improvement. 
With the latest \ac{BCI} technology, \citet{2110.07225} design a hand-free \ac{BCI}-based search system, which illustrates the possibility of using \ac{BCI} to replace keyboard and mouse in real life.
Therefore, using the brain signals for a proactive \ac{IR} system is promising and attracts much attention.

In the above work, the series of studies conducted by~\citet{moshfeghi2013understanding,moshfeghi2016understanding,moshfeghi2019towards} are most relevant to us.
The main differences between our paper and those \ac{fMRI} papers are: 
(1) Our research scenario is text-based reading comprehension while the \ac{fMRI} papers focuse on image-based relevance judgment.
Our findings uncover the brain activities when people locate key information during the reading and answer-seeking process. 
(2) We obtain different findings due to the high temporal resolution of \ac{EEG} devices, e.g., the finding in cognitive loading with N100-P200 and the interesting P600 phenomenon caused by semantic-thematic anomalous. 
(3) We conduct \ac{EEG}-based models to demonstrate the possibility to detect users’ reading status in real-time. 
\ac{EEG} is more portable than \ac{fMRI}, and therefore it is more meaningful to utilize \ac{EEG} signals for real-life classification tasks.
Results show the feasibility of constructing a better human-computer interaction system with brain signals.

%% file: 3-user_study.tex
\section{user study}
 
In the user study, participants are recruited to perform several reading comprehension tasks. 
Each trial includes a factoid question and the following sentence with graded relevance. 
Under a controlled user study setup in the prevention of potentially confusing effects, EEG data is recorded during the reading process. 
Open source of our code and dataset is in https://github.com/YeZiyi1998/UERCM.

\subsection{Participants}
We recruit 21 college students aged from 18 to 27~(M~\footnote{Mean value.} = 22.10, SD~\footnote{Standard deviation.} = 2.07).
Among them, there are 11 males and 10 females, who mainly major in computer science, physics, arts, and engineering. 
It takes about two hours to complete the whole task for each participant, including 40 minutes for preparation. 
Each participant is paid US\$30 after they complete all the tasks.

\subsection{Task preparation}

\subsubsection{Dataset}
\label{methdology:Experiment:Dataset}
For our user study, we first sample real-world questions from the WebQA~\cite{li2016dataset}, a factoid Q/A dataset, whose questions are open-domain with a close-ended answer.
We use this dataset for the following reasons: (1)~It is one of the largest Chinese Q/A datasets. (2)~It provides human annotation for correct answers and corresponding evidence.

More precisely, we manually sample 155 questions that cover topics including science, history, sports, and art. 
We generate three sentences for each question from this dataset and manually annotate each sentence with a relevance label.
Specifically, we select the ground truth sentence, the top sentence retrieved by BM25 but doesn't contain answer spans, and a randomly selected sentence as candidate sentences of \textit{perfectly relevant}, \textit{relevant}, and \textit{irrelevant}, respectively.
Further annotation is applied to verity and correct their relevance labels in Section~\ref{user study:Dataset:Annoataion} with the definitions of the relevance levels given in Section~\ref{The definitions of the relevance levels}.
 
Then, some of the sentences are manually refined to reduce the length and resolve grammar problems. 
Finally, the average question length is 8.7~(SD = 4.0), the average sentence length is 9.8~(SD = 3.0). 
Examples of sentences with different relevance levels are provided in Table \ref{tab:dataset_sample}. 

Following the above steps, we obtain a dataset consists of 155 questions and 465 sentences~(each question has three corresponding sentences). 
During the user study, participants will see a sentence randomly sampled from the three candidates for a given question.    

\begin{table}[t]
\caption{Example of user study tasks. The blod fonts and underlines indicate the answer words and the semantic-related words, respectively.}
\label{tab:dataset_sample}
 	\vspace{-3mm}
\renewcommand\arraystretch{1.3}
\begin{tabular}{ll}
\hlinew{0.8pt}
\textbf{Question}  & \begin{tabular}[c]{@{}l@{}}What is the largest mammal in the world?\end{tabular}    \\ \hline
\begin{tabular}[c]{@{}l@{}}\textbf{Perfectly} \\ \textbf{relevant} \end{tabular}& \begin{tabular}[c]{@{}l@{}}\textbf{The} \textbf{blue} \textbf{whale} is \underline{the} \underline{largest} \underline{animal} \underline{in} \underline{the} \underline{world}, \\ reaching an adult volume of 33 meters.\end{tabular}                    \\ \hline
\textbf{Relevant}           & \begin{tabular}[c]{@{}l@{}}\underline{The} \underline{largest} \underline{animal} \underline{in} \underline{the} \underline{world} in terms of supe-\\-rficial area is the Arctic chardonnay jellyfish.\end{tabular}   \\ \hline
\textbf{Irrelevant}         & \begin{tabular}[c]{@{}l@{}}It is estimated that there are about 10 billion capi-\\-llaries in human body.\end{tabular} \\ \hline 
\end{tabular}

\end{table}

\begin{figure*}[t]
\vspace{-3mm}
  \centering
  \includegraphics[width=1\linewidth]{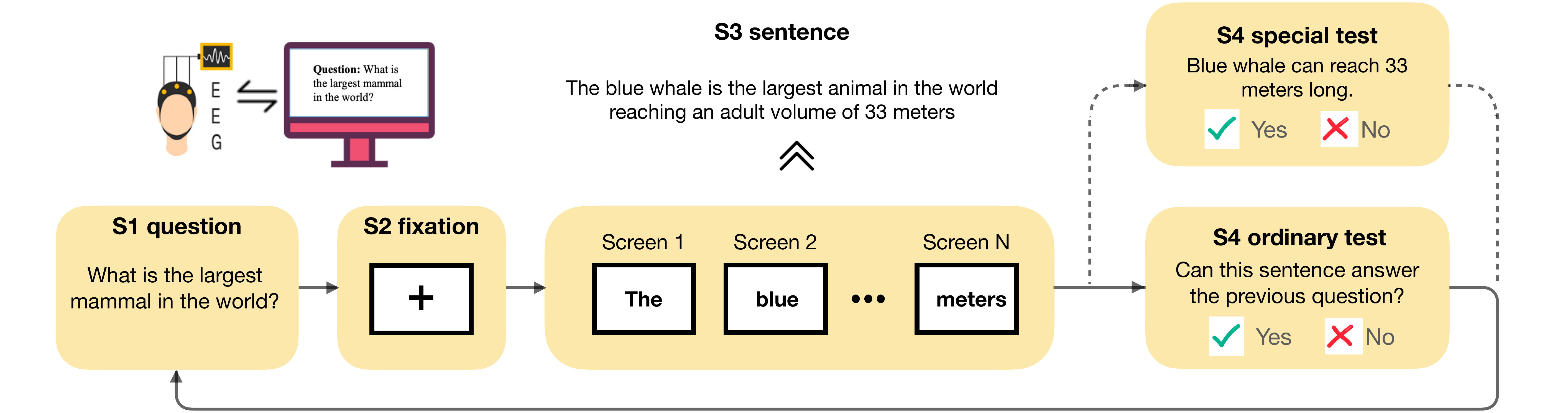}
  \caption{Structure of the main task. 
  First, a question is presented on the screen, and the participants can press the space key to skip after reading. 
  Then, a fixation cross and the words in the sentence are presented automatically in temporal sequence. 
  Third, an ordinary test or a special test is presented, and the participants should press a key to answer. 
  } 
  \label{fig:main task}
  \Description[]{}
  \vspace{-2mm}
\end{figure*}

\subsubsection{Annotation}
\label{user study:Dataset:Annoataion}
After constructing the reading comprehension dataset, we recruit three external assessors to annotate sentence-level graded relevance, identify answer words, and annotate semantic-related words.  
Examples of annotation data are provided in Table~\ref{tab:dataset_sample}. 
As a classification task for each word and sentence, the Fleiss's kappa between three annotators is 0.9542~(almost perfect agreement) for sentence-level relevance evaluation, 0.9343~(almost perfect agreement) for answer word identification, and 0.7848~(substantial agreement) for semantic-related words identification. 

\subsection{Procedure}
\label{Procedure}
This user study adheres to the ethical procedures for the protection of human participants in research and is approved by the ethics committee oof the School of Psychology at Tsinghua University.
The procedure of the user study, which consists of 6 stages, is detailed in the following.

\paragraph{Stage 1-4} 
In the beginning, Participants fill in an entry questionnaire to report demographic information and sign an informed consent about security and privacy protection. 
Then they read user study instructions about the main procedure of the user study. 
Prior to the main task, participants undergo a training step with five questions, which resembles the main task. 
The training step ensures that participants are familiar with the procedure of the main task.  
 
\paragraph{Stage 5}
Figure~\ref{fig:main task} illustrates the procedure of each trial in the main task. 
The main task contains 150 trials in total and is divided into six groups, each containing 25 trials.
The trials follow the same order of steps, i.e., S1 to S4 shown in Figure~\ref{fig:main task}: 
(S1)~Participants view a factoid question randomly selected from the dataset. 
Once they fully understand the question, they can press the space key and enter the second step. 
(S2)~A fixation cross is presented on the screen center to catch participants' attention and indicate the location of the following sentence presentation. 
The fixation cross will be presented for 1,000 milliseconds. 
(S3)~A sentence randomly selected from three candidates will be presented word by word, and each word will be shown for 750 milliseconds. 
The sequential presentation of words is a typical approach applied in natural sentence processing ERP studies~\cite{kutas1980reading}. 
The setting of reading pace is based on previous studies about carry-over of stimulus-evoked~\cite{dimigen2011coregistration}.
(S4)~Participants take a binary decision test about the question and the sentence. 
Two kinds of tests are randomly given.
The ordinary test is ``Can this sentence answer the previous question?'' and the special test is a binary factual judgment involving the sentence. 
The ordinary test is to confirm that participants have read the question carefully and are able to judge the relationship between the given question and the sentence.
While the special test is to ensure that, even if the participants can make the judgment of ordinary test beforehand, they should read the total sentence as well.
Finally, after the participants press the key~(J-key refers to ``Yes'' and F-key refers to ``No'') to pass the test, the next trial starts.
For each group, the test accuracy is checked to ensure that the participants perform tasks carefully.
The EEG data is recorded during the whole process with predefined triggers to locate time points of different steps.  

\vspace{-1mm}
\paragraph{Stage 6}
After completing the main task, they should fill in a post-questionnaires about the familiarity of given questions.

\subsection{Pilot study}
\label{Pilot study}
Before carrying out our main user study, a pilot study is conducted with four people outside the 21 participants to ensure the EEG recording system and the user study procedure work well. 
Detailed feedback obtained from pilot study participants is used to adjust the user study parameter settings, including the font size, amount of trials, time of rest period, etc.
Besides, following previous work~\cite{jiang2020alternative}, we adopt the ordinary test and the special test to ensure the participants can perform the question answering tasks carefully.
To adjust the probability for each kind of test, we set the ratio of special tests within ${10\%, 20\%}$ and find that the ratio of 10\% can achieve an accuracy of over 90\% for the special tests, while the ratio of 20\% achieves no significant improvements in that accuracy. 
Therefore, the probability for each kind of test is determined: 90\% for the ordinary test and 10\% for the special test.

\input{3.2-erp_analysis.tex}

%% file: 3.2-erp_analysis.tex
\subsection{\ac{ERP} methods}
\label{user study:Introduction to ERP}
\ac{ERP} is voltage generated in the brain structures in response to specific events or stimuli~\cite{blackwood1990cognitive}.
It usually refers to the brief \ac{EEG} data epoch, which is less than 1,000 ms after the experimentally designed stimuli. 
ERP components are evoked amplitudes in different time windows, including N100, N400 (negative wave in 100 ms, 400 ms), and  P200, P600 (positive wave in 200 ms,  600ms). 
Previous studies have revealed that ERP components are associated with neural activity with respect to both sensory and cognitive processes. 
The average waveform change between ERP components is also widely studied, such as the change from N100 component to P200 component~\cite{raney1993monitoring}.
To delve into the understanding of human reading comprehension from a neuroscience perspective, we apply standard \ac{ERP} analysis methods, including data pre-precessing, the division of time window and \ac{ROI}, and statistical methods, which are elaborated in Section~\ref{ERP analysis methods}.

%% file: 4-statistical_analysis.tex
\section{statistical analysis}

\begin{table}[t]
\caption{Statistical significance differences in all time windows and its ROIs~(see in Section~\ref{ROI}) among answer words~(A), semantic-related words~(S), and ordinary words~(O). $*/**$ indicate statistically significance at a level of $p \textless 0.05/0.001$ respectively using the post-hoc pair-wise Bonferroni's tests and the repeated measures ANOVA test.}
\label{tab:significance}
\begin{tabular}{llll}
\hline
\textbf{Time window}       & \textbf{ROI}        & \textbf{Post-hoc test}                       & \textbf{ANOVA p} \\ \hline
\multirow{2}{*}{120-320ms} & frontal    & A\textgreater{}S*                     & $*$       \\ 
                         & parietal   & A\textgreater{}O*                     & $*$       \\ \hline
\multirow{3}{*}{320-520ms} & central    & A\textgreater{}S*,A\textgreater{}O**  & $**$      \\ 
                         & r-temporal & A\textgreater{}O**                    & $**$     \\ 
                         & parietal   & A\textgreater{}S*,A\textgreater{}O**  & $**$     \\ \hline
\multirow{3}{*}{520-750ms} & central    & A\textgreater{}S**, A\textgreater{}O** & $**$      \\ 
                         & l-temporal & A\textgreater{}S**, A\textgreater{}O*, S\textless{}O*  & $**$      \\  
                         & parietal   & A\textgreater{}S*, A\textgreater{}O**   & $**$     \\ \hline
\end{tabular}
\vspace{-3mm}
\end{table}

\subsection{Questionnaire and Behavioral Response}

A post-questionnaire is used to collect the users' perceived familiarity level on the topics of all the questions, with a five-point Likert scale~(\textit{Highly familiar}, \textit{Somewhat familiar}, \textit{Neither familiar nor unfamiliar}, \textit{Somewhat unfamiliar}, \textit{Totally unfamiliar}). 
About one-third of questions are reported familiar to the users~(\textit{Highly familiar}: 21.07\%, \textit{Somewhat familiar}: 16.85\%) and another one-third unfamiliar to the users~(\textit{Somewhat unfamiliar}: 26.9\%, \textit{Totally unfamiliar}: 3.78\%). The rest of them are reported to be \textit{Neither familiar nor unfamiliar}(31.4\%).
ERP analysis shows no significant difference in our study across different familiarity levels.
It reveals that no matter how familiar the user is, the reading process will evoke similar patterns in the brain.    

The behavioral responses are analyzed in terms of the accuracy rate and the reaction time of the binary decision test. 
The accuracy rate is 97.93\% for \textit{perfectly relevant}, 92.03\% for \textit{relevant}, and 89.98\% for \textit{irrelevant}, while the reaction time is 1.00s for \textit{perfectly relevant}, 1.29s for \textit{relevant}, and 1.39s for \textit{irrelevant}. 
These results indicate that behavioral responses are different accordingly, considering the graded relevance of sentences. 
Therefore, we can speculate that neurological factors exist behind these differences, which is essential to study. 

\subsection{ERP Components}
\label{statistical analysis:ERP Components}
\begin{figure}[t]
\vspace{-3mm}
  \centering
  \includegraphics[width=0.8\linewidth]{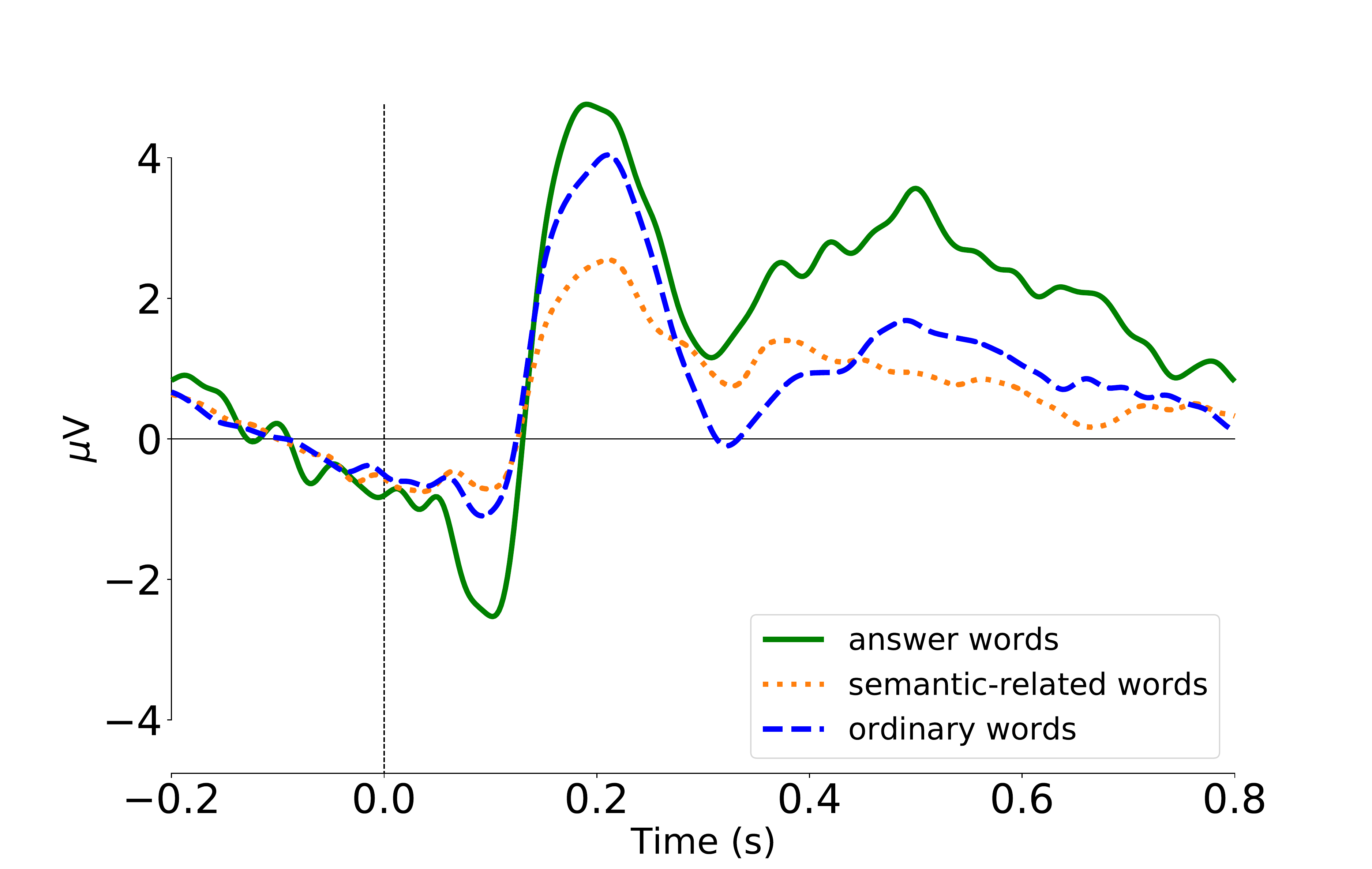}
  \caption{The grand average ERP waveforms in central~(Cz + FCz + C3 + C4 + FC3 + FC4) by word types.} 
  \label{fig:word3}
  \Description[]{}
  \vspace{-3mm}
\end{figure}

Significance levels of observed differences are reported in Table~\ref{tab:significance}.
Besides, Figure~\ref{fig:word3} provides the grand average ERP waveforms of words across different types~(answer words, semantic-related words, and ordinary words) in central. 
In detail, we have the following observations: 

\paragraph{120-320ms}
\label{120-320}
In the 120-320ms time window, the differences of P200 waveforms generated by different types of words are slightly significant~(p $<$ 0.05 in frontal and parietal).
And there exist highly significant differences in the N100-P200 amplitudes~(the average waveform change from N100 to P200) in frontal~(F[2, 40] = 19.51, p < 0.001), central~(F[2, 40] = 20.94, p < 0.001) and parietal~(F[2, 40] = 29.14, p < 0.001). 
The Bonferroni's test reveals that the N100-P200 amplitude of answer words is significantly higher than that of semantic-related words~(p < 0.001) and ordinary words~(p < 0.001).

Previous studies indicate that the lower cognitive loading in reading is associated with an increase in the N100-P200 amplitude~\cite{raney1993monitoring}. 
This increase in N100-P200 amplitude of answer words may suggest that cognitive resources are less demanded when users locate the answer.

\paragraph{320-520ms}
The grand-averaged N400 component waveforms in the 320-520ms time window after the word stimulus onset are examined, showing significant differences in central~(F[2,40] = 12.57, p < 0.001), r-temporal~(F[2,40] = 17.34, p < 0.001), and parietal~(F[2, 40] = 15.59, p < 0.001). 
The Bonferroni's test reveals that the mean negativity of answer words in N400 is significantly smaller than that of semantic-related words~(p < 0.05) and ordinary words~(p < 0.001). 
Besides, the mean negativity of the semantic-related words is significantly smaller than that of ordinary words~(p < 0.05) in electrodes T4 and T6.  

N400 is well-known to be associated with the message-level representation on the processing of upcoming words~\cite{kutas1984brain,hoeks2004seeing}. 
The higher ``expectedness'' of a word in the current semantic context usually leads to a smaller N400 negativity. 
Our statistical analysis suggests that the N400 negativity of answer words is smaller than that of the semantic-related words.
The N400 negativity of semantic-related words is again smaller than that of ordinary words.
The finding of ``expectedness'' is consistent with the previous finding of cognitive loading in Section~\ref{120-320} since words of higher ``expectedness'' may need a less cognitive resource. 
Additionally, our findings also imply that semantic-related words have higher ``expectedness'' than ordinary words.

\paragraph{520-750ms.}
The P600 waveforms evoked by the stimulus are grand-averaged in the 520-750ms time-window, which show significant effect in central~(F[2, 40] = 17.45, p < 0.001), l-temporal~(F[2, 40] = 15.87, p < 0.001), parietal~(F[2, 40] = 20.27, p < 0.001). 
The Bonferroni's test reveals that the mean positivity of answer words in P600 is significantly larger than that of semantic-related words~(p < 0.001) and ordinary words~(p < 0.001) in central.
Besides, the mean positivity of the semantic-related words is significantly smaller than that of ordinary words in l-temporal~(p<0.01). 
 
Recent studies reveal that P600 is associated with semantic-thematic anomalous~\cite{van2005erp} and inferential processing~\cite{burkhardt2006inferential}.
In \ac{IR} scenario, \citet{eugster2016natural} show relevant words would elicit higher P600 amplitudes. 
\citet{pinkosova2020cortical} indicate that the link between higher relevance and P600 amplitude might come from discourse memory in the brain.
In our study, sentences have no problem at the syntactic level after we check manually. 
Thus, we speculate that the differences among different contents may be caused by semantic-thematic anomalous and inferential processing.
Both of these aspects are also related to discourse memory, as \citet{pinkosova2020cortical} indicate. 

More specifically, it is interesting to find that P600 is the highest in answer words, followed by ordinary words, while lowest in semantic-related words, especially in l-temporal~(related to language recognition). 
For answer words, it is obvious that inferential processing is initiated in human's brain, causing significantly higher P600. 
Similarly, semantic-related words may also relate to inferential processing, but to a less extent. 
Both types of words have a minimal relationship with semantic-thematic anomalous since they are semantically correlated. 
Nevertheless, for ordinary words, semantic-thematic anomalous becomes dominant compared to semantic-related words since it is less helpful for semantic-thematic understanding.
Thus ordinary words result in a relatively high P600 amplitude.
Generally speaking, it is most likely that semantic-related words would cost relatively low discourse memory. 
However, this interesting phenomenon needs further exploration to uncover the underlying neural mechanism. 

\subsection{Discussions}

In conclusion, our findings take an important step towards unraveling the nature of reading comprehension and, in turn, enlighten search systems that are more proactive and human-friendly.
On the one hand, the ERP analysis across time windows shows that neural differences exist between processing key information and ordinary information during reading comprehension~(addressing \textbf{RQ1}). 
On the other hand, we believe that various cognitive activities, e.g., cognitive loading, semantic-thematic understanding, inferential processing, underpin these neural responses~(summarized in Figure~\ref{fig:topmap}).
Different from previous studies using eye-gazing data~\cite{li2018understanding}, our findings are built on a deeper cognitive level involving how human process text information. 
These cognitive differences can help us understand reading comprehension process and illustrate several insights for \ac{IR} tasks~(addressing \textbf{RQ2}):

 \begin{figure}[t]
  \centering
  \includegraphics[width=1\linewidth]{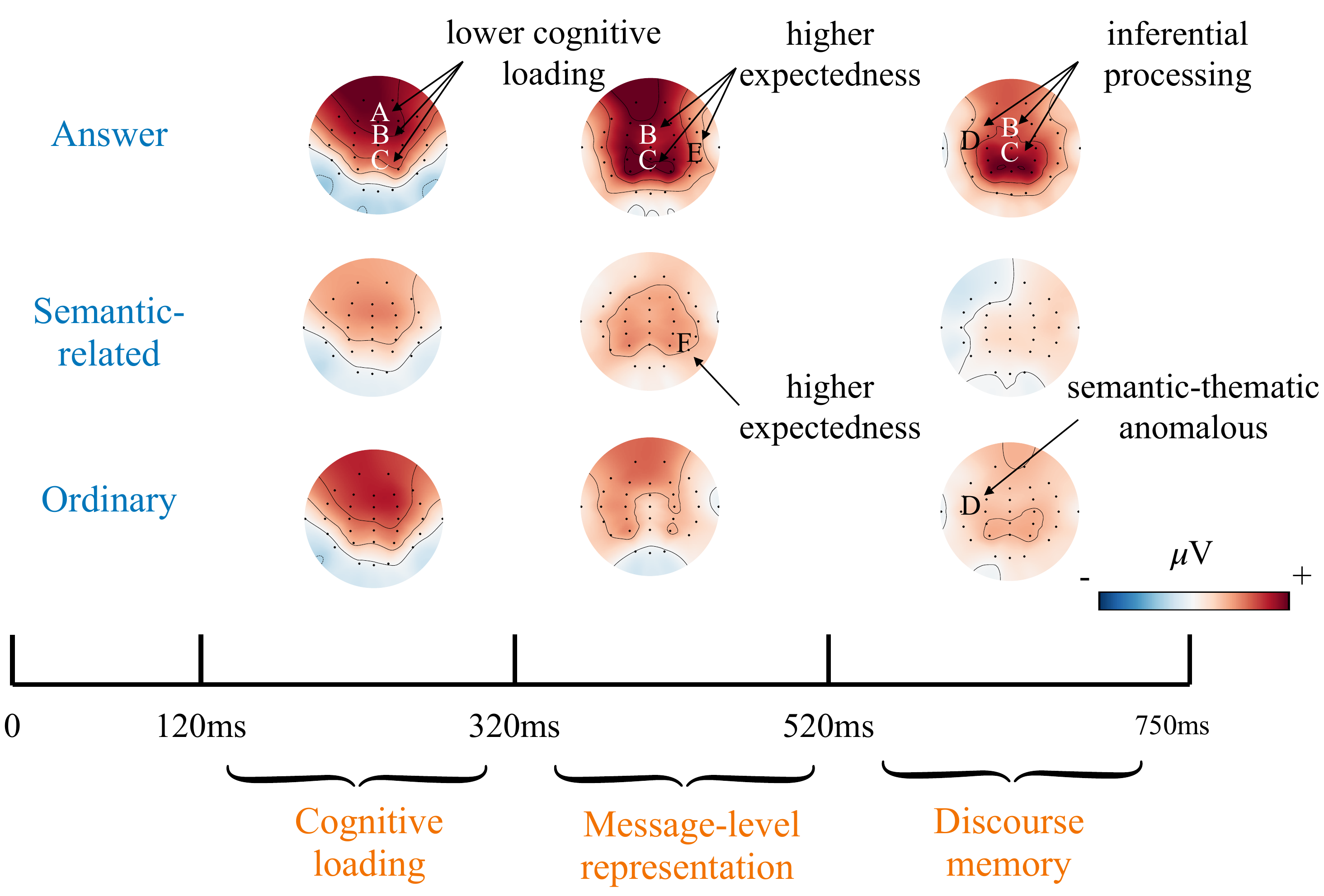}
  \setlength{\abovecaptionskip}{-1.5mm}
  \caption{Condition-wise topographies averaged across time windows for three word types and possible potential mental phenomena. A, B, C, D, E, and F refer to frontal, central, parietal, l-temporal, r-temporal, and electrode T6, respectively. The lower and upper bounds are $\pm$5 $\mu$V for 120- 320 ms time window, and $\pm$3 $\mu$V for other time windows.} 
  \vspace{-3.3mm}
  \label{fig:topmap}
  \Description[]{}
\end{figure}

(1)~\textbf{Insights for document ranking.}
The finding in the N100-P200 amplitudes illustrates that cognitive resources are less demanded when participants locate the answer.
The demand for cognitive resources, namely as cognitive capacity~\cite{jiang2021neural}, affects the user experience with external system.  
Besides, \citet{jiang2021neural} suggest that reduced cognitive capacity can cause impaired detection accuracy, which is related to misunderstanding in reading comprehension scenarios.
Therefore, owing to the decrease of cognitive capacity during reading answer contents, we think that easily accessible contents of potential answers are important.
A better form of the document structure is the combination of concise and brief key information contents and exhaustive supplementary contents.
In practice, the search engine should consider fine-grained document structure, especially the position and display styles of potential answer, when constructing ranking models. 

(2)~\textbf{Insight for the construction of result snippets.}
 When users locate the answer, we speculate that they will switch their extra cognitive resources into other neurological functions~(e.g., the expansion of working memory capacity for information recall and management).
The findings in the P600 effect of answer contents, which implicate the occurrence of inferential processing in human brain, verify our speculation.
Besides, we also find that semantically related content requires few inferential processing functions, thus its P600 effect is even smaller than that of ordinary contents.  
In the current search interface design, we usually find the result snippets on the \ac{SERP} contain much semantic related content but omit ordinary information.
Although providing much semantic related content makes a search result attractive~(higher expectedness as illustrated in our \ac{ERP} analysis), it may cause unsatisfactory after clicking the results in certain situations. 
Our results suggest that search engines should consider extracting snippets fairly by considering not only semantic factors but also factors related to whether the content can provide evidence and background for a better understanding.

(3)~\textbf{Insight for \ac{BCI}-enhanced search system.}
As \ac{BCI} devices becoming low-cost and portable~\footnote{https://the-unwinder.com/reviews/best-eeg-headset/}, researchers have suggested a revolution of online \acp{BCI} in the near future~\cite{liu2010online}, which can be applied in online education, internet surfing, and search.
Especially in search scenarios, researchers have realized a free-hand system for search with \ac{BCI}~\cite{2110.07225}.
Since reading comprehension is a common task in these scenarios, utilizing \ac{BCI} to acquire a better understanding of reading status is possible, and we believe it will benefit human computer interactions.
For instance, with \ac{BCI}, search engines can understand what content satisfys the users and further provide more helpful information, especially in circumstances where user intents are ambiguous.
With the finding of these detectable differences, we are encouraged to explore the effectiveness of using brain signals as implicit feedback for reading comprehension tasks, which is elaborated in Section~\ref{experiments and discussions}.

%% file: 5-experiments_and_discussions.tex
\section{experiments and discussions}
\label{experiments and discussions}
To explore the reading process, we conduct two experimental tasks, i.e., answer sentence classification and answer extraction, based on the EEG data collected in our user study.
These tasks are crucial in the study of machine reading comprehension~\cite{yao2013answer,severyn2013automatic,li2016dataset} and \ac{IR}~\cite{li2018understanding}.
Note that we aim to demonstrate the effectiveness and interpretability of \ac{EEG} signals as implicit feedback, investigation on constructing more sophisticated models considering brain signals and other interactive features is left as future work.
 
\subsection{Models}
Given word-level \ac{EEG} features~(details can be seen in Section~\ref{Features}), the answer extraction task is a binary classification problem to estimate the possibility of a word being the answer.
And the answer sentence classification task is a classification problem of estimating the probability that a sentence being \textit{perfectly relevant}.

To resolve these problems with a unified framework and show the effective of \ac{EEG} signals, we propose a \textbf{U}nified framework for \textbf{E}EG-based \textbf{R}eading \textbf{C}omprehension \textbf{M}odeling~(UERCM).
The framework provides a common structure for these two tasks, which considers learnable positional encoding and attention mechanism to capture the local interactions of \ac{EEG} features in a sentence.
Although attention mechanism has been widely applied in various \ac{NLP} tasks~\cite{vaswani2017attention}, few studies have shown its effectiveness in \ac{EEG}-based scenarios.
Thus we aim to illustrate the effectiveness of brain signals utilizing attention mechanism, the design of more sophisticated model is left as future work.

\begin{figure}[t]
  \centering
  \includegraphics[width=0.65\linewidth]{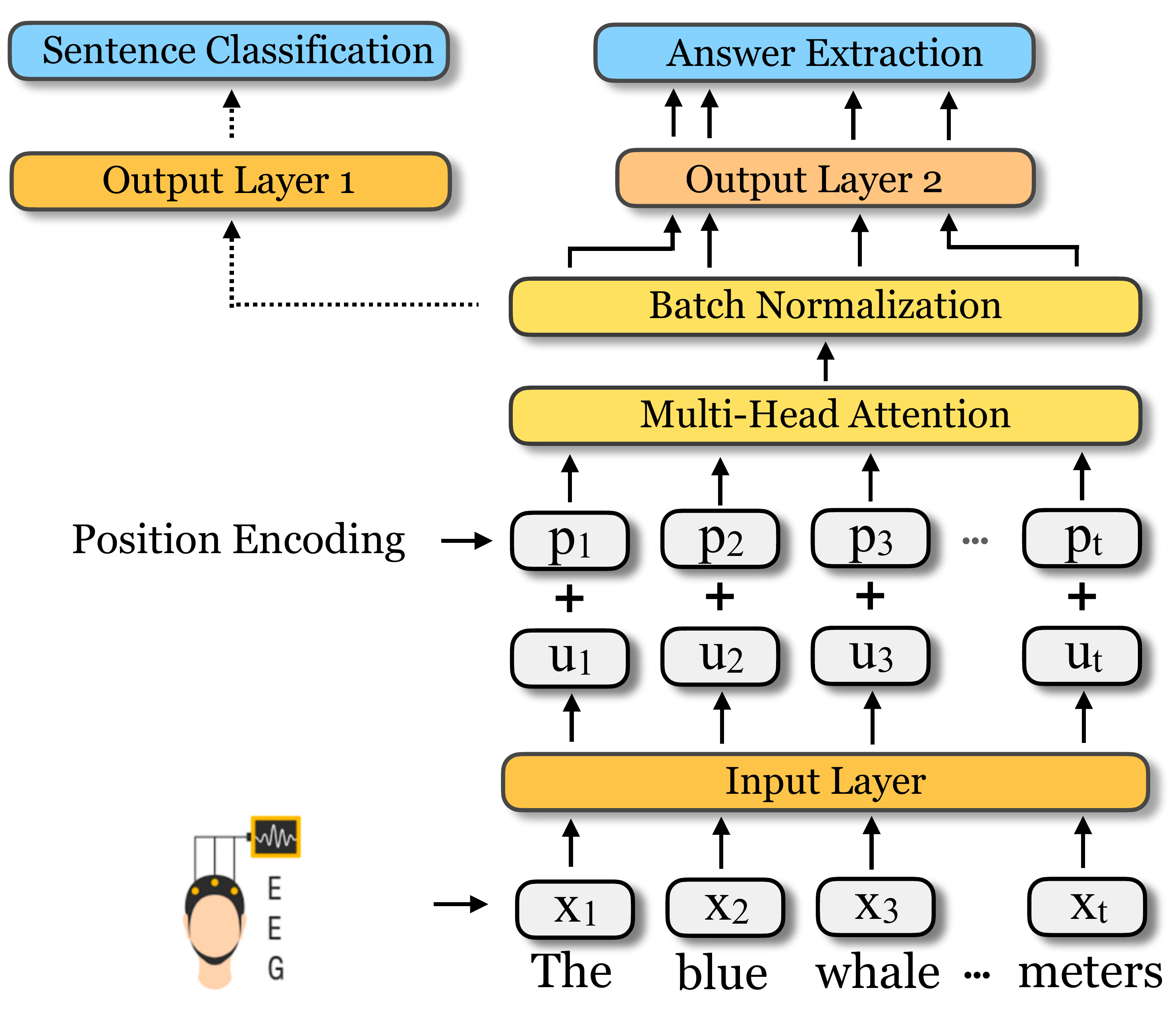}
  \caption{The proposed framework of UERCM.} 
  \label{fig:model}
  \vspace{-3mm}
\end{figure}

The proposed framework is presented in Figure~\ref{fig:model}.
For a specific word-level \ac{EEG} sequence $X \in \mathbb{R}^{t \times d} = [x_1,x_2,...,x_t]$, where $t$ is the sentence length and $d$ is the length of word-level \ac{EEG} features, we first apply an input layer to linearly project it onto a $h$-dimensional vector space, where $h$ is the hidden dimension of the transformer model sequence element representations:
$$
U=W_{h}X+b_{h}
$$
where $W_{h} \in \mathbb{R}^{d \times h}$ and $b_{h}\in \mathbb{R}^{t \times h}$ are learnable parameters and $U \in \mathbb{R}^{t \times h}$ is the hidden vectors, which is later adopted as the input for the multi-head attention layer.
After that, we add position encodings $P \in \mathbb{R}^{t \times h} = [p_1,p_2,...,p_n]$ into the vector $U$ and acquire $U'\in \mathbb{R}^{t \times h}=U+P $.  
Instead of sinusoidal encodings~\cite{vaswani2017attention}, we apply learnable positional encodings since they perform better.
Then, we apply a multi-head attention layer to calculate the local-interacted sequence:
$$
Z=MultiHead(U',U',U')
$$
where $Z \in \mathbb{R}^{t \times h} $ is the output vector.
Next, we apply a batch normalization layer to accelerate the training procedure and get $Z'=BatchNormalization(Z)$.
\citet{vaswani2017attention} suggest using layer normalization after the multi-head attention layer, which leads to performance gains over batch normalization in various \ac{NLP} tasks.
In spite of this, we find batch normalization performs better than layer normalization in our tasks.
We suggest the reason is batch normalization can mitigate the effect of instability in \ac{EEG} features, an issue that does not arise in pre-trained \ac{NLP} word embeddings.
In addition to the design of batch normalization, another difference of our framework exists in the number of attention layers~(when compared to the application of attention mechanism in \ac{NLP} tasks).
In stead of utilizing several attention layers for better presentation ability, we simply use one attention layer to avoid the increasement of parameters, which performs better in our experiment.
Since we aim to verify the effectiveness of attention mechanisms and sequence modeling in the novel EEG-based reading comprehension tasks, the discussion of the parameters sensitivity and in-depth comparison between  Transformers~\cite{vaswani2017attention} in \ac{NLP} is left as future work.

After that, given the representation $Z$, we adopt different strategies for two tasks to aggregate them to get the prediction of sentence classification~($\hat{y}_s \in \mathbb{R}^{1}$) and answer extraction~($\hat{Y}_o \in \mathbb{R}^{t} =[\hat{y}_{o,1},\hat{y}_{o,2},...,\hat{y}_{o,t}]$), respectively:
$$
\hat{y}_s=softmax(W_sReLU(Concat(z_1,z_2,...,z_t))+b_s)
$$
$$
\hat{y}_{o,i}=softmax(W_oReLU(z_i)+b_o),i=1,2,...,t
$$  
where $W_s \in \mathbb{R}^{th \times 1} $, $b_s \in \mathbb{R}^{1}$, $W_o \in \mathbb{R}^{h \times 1} $, and $b_o \in \mathbb{R}^{1}$ are parameters for the linear output layer.

Finally, we adopt the cross-entropy function as learning objective, the loss $L_s$~(in answer sentence classification task) and $L_o$~(in answer extraction task) for a sample sentence are:
$$
L_s = - y_slog\hat{y}_s+(1-y_s)log((1-\hat{y}_s))
$$
$$
L_o = - \sum_{i}{(y_{o,i}log\hat{y}_{o,i}+(1-y_{o,i})log((1-\hat{y}_{o,i})))}
$$
where $y_s \in \mathbb{R}^{1}$, $y_{o,i} \in \mathbb{R}^{1}$ are ground truth values of sentence label and word label, respectively. 
 
For training procedures, we optimize two tasks (1)independently and (2)~jointly except for the last output layer.
Their performance are similar, and thus, we only report the experiment results with independent training procedures.
For baselines, we adopt an untrained model, \ac{SVM}, \ac{MLP}, \ac{GBDT}, and \ac{RNN}~(with a particular \ac{CRF} module for answer extraction task).
The details are provided in Section~\ref{Experimental Settings}. 
We do not apply other sophisticated models as baselines because: (1)~there is no specially designed neural model for this task previously and (2)~we aim to show the effectiveness of brain signals and provide a unified solution, the comparisons with more models are left as future work.

To verify the effective in different settings, we perform two data splitting strategies to deal with unseen questions and unseen users: \ac{CVOT} and \ac{LOPO}, respectively. 
As for evaluation metrics, keeping with prior work~\cite{rodrigo2014evaluating,yao2013answer}, we use \ac{AUC} for both answer extraction and answer sentence classification and \ac{MAP} for answer sentence classification since it can also be treated as a ranking problem.
And we calculate the margin~($\Delta AUC$ and $\Delta MAP$) of other models compared to the untrained model to illustrate the effectiveness of brain signals.
More detailed experimental settings~(i.e., parameter setups and dataset splitting strategies) are elaborated in Section~\ref{Experimental Settings}.

\subsection{Results and Discussions}

\begin{table*}[t!]
\caption{Experimental result of answer extraction and answer sentence classification. $*$/$\dagger$ indicates difference compared to the untrained model~(see in Section~\ref{Experimental Settings}) and UERCM is significant with $p$-value $\textless 0.05$, respectively.  
}
\label{tab:result}
\setlength{\tabcolsep}{3mm}{

\begin{tabular}{@{}l!{\color{lightgray}\vrule}cc!{\color{lightgray}\vrule}cccc}
\toprule
\multirow{2}{*}{\textbf{Model}}& \multicolumn{2}{c!{\color{lightgray}\vrule}}{\textbf{Answer Extraction}}   & &\multicolumn{2}{c}{\textbf{Answer Sentence Classification}} & \\
 & {$\Delta \ac{AUC}_{CVOT}$}  & {$\Delta \ac{AUC}_{LOPO}$}          & {$\Delta \ac{AUC}_{CVOT}$}    &  {$\Delta \ac{MAP}_{CVOT}$} &  {$\Delta \ac{AUC}_{LOPO}$} &  {$\Delta \ac{MAP}_{LOPO}$} \\ \midrule   
SVM    & {$0.072^{*\dagger}$}  &  {$0.069^{*\dagger}$}  & {$0.092^{*\dagger}$} & {$0.065^{*\dagger}$} & {$0.103^{*\dagger}$} & {$0.078^{*\dagger}$}\\ 
MLP   & {$0.079^{*\dagger}$} &   {$0.084^{*\dagger}$}   & {$0.141^{*\dagger}$} & {$0.077^{*\dagger}$} & {$0.122^{*\dagger}$} & {$0.086^{*\dagger}$}\\
GBDT    & {$0.086^{*\dagger}$} & {$0.077^{*\dagger}$} & {$0.097^{*\dagger}$} & {$0.079^{*\dagger}$} & {$0.125^{*\dagger}$} & {$0.074^{*\dagger}$}\\
RNN~(+CRF) & {$0.146^{*\dagger}$} & {$0.151^{*}$ }  & {$0.132^{*}$  } & {$0.089^{*\dagger}$} & {$0.165^{*\dagger}$} & {$0.101^{*\dagger}$}\\
UERCM     & {$\textbf{0.152}^{*}$ } & {$\textbf{0.157}^{*}$ } & {$\textbf{0.173}^{*}$ } & {$\textbf{0.147}^{*}$ } & {$\textbf{0.236}^{*}$ } & {$\textbf{0.179}^{*}$ } \\
\bottomrule

\end{tabular}}
\end{table*}

Table~\ref{tab:result} presents the experimental results of the answer extraction task and answer sentence classification task in two dataset splitting strategies(\ac{CVOT} and \ac{LOPO}, see in Section~\ref{Experimental Settings}), respectively.
Generally, it can be seen that all the models based on the EEG features are significantly better than the untrained model. 
These results demonstrate the feasibility of using EEG data to locate answer words and monitor users' answer-seeking process.
Besides, UERCM achieves the best performance, especially it improve a large margin on the answer sentence task compared to all baselines in different dataset splitting strategies.
This suggests the strategies of attention mechanism and the sequence modeling of brain signals enable UERCM to outperform both machine learning baselines~(\ac{SVM} and \ac{GBDT}) and neural baselines~(\ac{MLP} and \ac{RNN}~(+\ac{CRF})) substantially.

Now we delve into the performance of different models in these two tasks, respectively.
We observe that:

(1)~For answer extraction, the baseline models of SVM, MLP, and GBDT perform significantly worse than UERCM.
The reason may be that they treat the task as a binary decision for each word while missing the sequence information.
Conversely, models adopting sequence modeling strategies, i.e., RNN~(+CRF) with conditional probability estimations and UERCM with local interactions, substantially outperform other baselines.
Although UERCM does not perform significantly better RNN~(+CRF) in this task, we suggest it would be a better solution since it can realize parallel computation while RNN can't due to its iteration nature.
For a real-time \ac{BCI} equipment, UERCM can accelerate the inference process and save time for later calculation utilizing this implicit feedback.

(2)~For answer sentence classification, UERCM leads to a significant improvement and outperforms other baselines significantly on $\Delta MAP$ in both data splitting strategies.
Especially, we find RNN~(+CRF) performs significantly worse than UERCM, although it considers sequence modeling as well. 
This phenomenon may be caused by the specificity of brain signals. 
Brain signals commonly contain fluctuations such as blinks and heartbeats.
Although we apply standard preprocessing methods, the data quality is still unstable for brain signals related to some words.
For the RNN model, the performance suffers from the bad signals during the iteration process.
But our UERCM is stable since it can automatically mitigate the effect of bad signals and extract valuable information from other data in the sequence with the local interaction.

\paragraph{Answer to \textbf{RQ3}}
Experimental results in answer extraction and answer sentence classification tasks suggest that \ac{EEG} signals can be leveraged to classify answer sentences and extract the answer words. 
Besides, our proposed framework UERCM outperforms other baselines in both \ac{CVOT} and \ac{LOPO} settings.

%% file: 6-conclusion.tex
\section{conclusion} 
In this paper, we have studied brain activities under a reading comprehension scenario. 
We have investigated the cognitive responses when users locate different text contents, including answer span contents, semantic-related contents, and other ordinary contents.
Our analysis contributes to a better understanding of reading comprehension.
Major findings and insights for \ac{IR} include:
(1)~There are detectable differences in neural activities between contents that can satisfy the information need and contents that can not. 
These differences are related to cognitive loading, ``expectedness'', inferential processing, and other aspects.  
(2)~The findings in N100-P200 waveforms, which are related to cognitive 
capacity, suggest that the ranking model construction should consider fine-grained document structure to reduce cognitive load and avoid misunderstanding.
(3)~Inferential processing is crucial in human reading comprehension, which is expressed as the P600 effect in our analysis. Thus search engines should consider factors beyond semantic similarity when extracting the snippets on \ac{SERP} for better user experience.  

As \ac{EEG} devices becoming low-cost and portable, researchers have suggested applying \ac{BCI} for scenarios including education, internet surfing, and search.
Therefore, we think that \ac{BCI} can be useful to detect users' reading and answer-seeking state for better human-computer interactions. 
To address this issue, we propose a novel framework UERCM, which can effectively classify answer sentences and extract answers during reading comprehension.
To our best knowledge, this is the first work utilizing brain signals for these tasks.
And the experimental results show that brain signals can be used as valuable implicit feedback during reading comprehension.

Our study is limited to a lab-based sentence-level reading comprehension scenario under our experimental paradigm.
The limitations may guide future works such as:  
(1)~Although portable \ac{EEG} devices contain more noise than lab-based devices, we believe the technology will have a revolution in the near future. 
Thus it is promising to collect brain signals in real-life reading comprehension tasks and consider other interaction features~(e.g., query generation, mouse movement) to construct \ac{BCI}-enhanced information system.
(2)~Our cognitive findings with human reading comprehension provide several insights to facilitate the ranking models construction and search interface design.
Empirical studies with real-world search engines are meant to verify our findings beyond the neuroscience perspective.
(3)~We propose a novel framework UERCM in our work as a first step to detect human reading comprehension with brain signals.
It is interesting to explore other practical \ac{IR} tasks with \ac{BCI} and design more sophisticated models for better performance.

%% file: 7-acknowledgements.tex
\section{Acknowledgement}
This work is supported by the Natural Science Foundation of China~(Grant No. 61732008), Beijing Academy of Artificial Intelligence~(BAAI), and Tsinghua University Guoqiang Research Institute.

%% file: 7-appendix.tex
\section{SUPPLEMENTARY MATERIAL}

\subsection{The definitions of the relevance levels}
\label{The definitions of the relevance levels}
The definitions of the relevance levels are:

\begin{itemize}
	\item \textit{Perfectly relevant}: The sentence is dedicated to the question so we can get the exact answer to the question. It is worth being a top result in a search engine.
	\item \textit{Relevant}: The sentence provides some information relevant to the question. It is semantic relevant, but its contribution to solving the question may be minimal.
	\item \textit{Irrelevant}: The sentence does not provide any useful information about the question, and it is semantic irrelevant.
\end{itemize}

These definitions are modified from the definitions in TREC 2019 deep learning track~\cite{craswell2020overview} with four relevance levels.
We merge the relevance level of \textit{highly relevant} and \textit{relevant} into \textit{relevant} in our definition to simplify task settings.

\subsection{Apparatus}
Our study uses a laptop computer with a 17-inch monitor with a resolution of 1,600×900. 
A 40 electrodes Scan NuAmps Express system~(Compumedics Ltd., VIC, Australia) and a 37-channel Quik-Cap~(Compumedical NeuroScan) are deployed to capture the participants’ EEG data. 

\subsection{\ac{ERP} analysis methods}
\label{ERP analysis methods}
\subsubsection{Data pre-processing}
\ac{EEG} data is pre-processed according to standard procedures.
First, the EEG data is re-referencing to average mastoids~(A1 and A2).
Second, baseline correlation for each channel is applied to remove fluctuations in the signal. 
Third, the EEG data was filtered in the frequency range of 0.5–30.0 Hz to preserve the \ac{EEG} frequency band.  
Fourth, we perform a parametric noise covariance model~\cite{huizenga2002spatiotemporal} to remove  components associated with ocular, cardiac, and muscular artifacts. 
Moreover, epochs~(brief \ac{EEG} segment, 1000ms in our experimental settings) with an absolute maximum voltage over the threshold 100 $\mu$V are marked as bad. 
Fifth, the \ac{EEG} data is down-sample to 500 Hz for the following analysis. 
Finally, interested epochs are extracted according to the triggers~(time points to locate interested \ac{EEG} data, see in Section~\ref{Procedure}), and baseline corrected using the pre-stimulus period -200-0 ms. 
\acp{ERP} are averaged across the same type of words for further analysis.

\subsubsection{Time window}
To distinguish \ac{ERP} components, time windows are split in our analysis.
\citet{lehmann1980reference} propose a method to determine components of evoked scalp potentials in terms of times of occurrence~(latency) and location on the scalp~(topography), which is one of the most established measures in \ac{ERP} mapping. 
The Global Field Power~(GFP) is calculated between 0-750 ms, and we determine time segments according to the power distribution. 
As a result, the determined time segments are 60-120 ms for N100, 120-320 ms for P200 component, 320-520 ms for N400 components, and 520-750 ms for P600 component, respectively. 
As an early component, N100 is a pre-attentive potential which does not involve semantic understanding of textual content~\cite{luck2000event,vogel2000visual}.
Hence we only discuss findings in P200, N400, and P600 components in this article.

\subsubsection{\ac{ROI}}
\label{ROI}
Different brain areas have different functions, e.g., parietal is associated with logistics and mathematical thinking.
In the field of \ac{IR}, frontal, parietal, and r-temporal are implied to be related to relevance judgments~\cite{moshfeghi2013understanding}.
In that regard, it is necessary to identify \ac{ROI}.
In particular, a permutation T-test is applied on sensor data in a fixed time window for each ERP component. 
Then ROI is identified based on the active sensors as well as their spatial distribution.
Electrodes are assigned to seven brain areas according to their spatial distribution: prefrontal, frontal, central, parietal, l-temporal, r-temporal, and occipital. 
 The selected \ac{ROI}s for each time window are shown in \ref{tab:significance}.

\subsubsection{Statistical Methods}

In order to test the difference of ERP components between different types of words, we applied repeated measures ANOVA. 
The independent variable is the three types of words: answer words, semantic-related words, and ordinary words. 
Examples of different types of words can be found in Table~\ref{tab:dataset_sample}.
The dependent variable is the mean signal in a given time window and ROI. 
We have tried to combine the effect of sentence relevance and find the results of \ac{ERP} analysis are similar.
Thus they are not reported in our study.
Before the multi-group comparsion, Shapiro-Wilk's test is applied to check the normality of data. 
To check the feasibility of repeated measures ANOVA, each condition's sphericity assumption is verified using the Mauchly’s test.
Then the Greenhouse-Geisser method is applied when the sphericity is not met. 
Finally, we apply post hoc Bonferroni tests to conduct pair-wise comparisons between groups. 

\subsection{Features}
\label{Features}
Previous works in EEG feature engineering contain two major types of EEG features, i.e., \acp{FBF} and \acp{ERPF}.
On the one hand, \acp{FBF} capture frequency information during the whole time window.
Frequency information in different bands is associated with attentiveness~(delta~\cite{harmony1996eeg} and beta~\cite{klimesch1999eeg}), cognitive performance~(theta and alpha~\cite{klimesch1999eeg}), and semantic violation~(gamma~\cite{penolazzi2009gamma}). 
Previous works have shown the effectiveness of \acp{FBF} for relevance prediction~\cite{gwizdka2017temporal,gwizdka2018inferring}.
On the other hand, \acp{ERPF} capture the time domain information within a specific short time window when users receive a stimulation.
Previous works in \ac{BCI} systems have shown the effectiveness of ERP components, such as N170 and P300, in terms of online target detection~\cite{zhang2012novel}.
In the field of \ac{IR}, \acp{ERPF} are shown to be associated with relevance judgments~\cite{jacucci2019integrating,eugster2014predicting} and decision making in information seeking~\cite{frey2013decision}.
Analyses in Section~\ref{statistical analysis:ERP Components} also show the potential correlation between \acp{ERP} and the information seeking process during reading comprehension. 
For the above reason, \acp{FBF} and \acp{ERPF} are extracted in our study.

Concretely speaking, we select EEG features from three brain regions~(central, r-temporal, and parietal).
The reason is that these regions have significant differences in cognitive responses across different types of words, as shown in Section~\ref{statistical analysis:ERP Components}.
For \acp{FBF}, average band power and differential entropy are calculated from the frequency bands of delta~(0.5-4Hz), theta~(4-8Hz), alpha~(8-13Hz), and beta~(13-30Hz).
For \acp{ERPF}, five time points are evenly sampled from P200~(120-320ms), N400~(320-520ms), and P600~(520-750ms), respectively.
As a result, the representation for each word is a 69-dimensional vector~(2$*$4 \acp{FBF}$*$3 regions,5$*$3\acp{ERPF}$*$3 regions) that contains information from EEG data.
To analyze the feature importance in different \ac{ERP} components and \ac{ROI}, as discussed in Section~\ref{statistical analysis:ERP Components}, we present the \ac{SHAP}~\cite{lundberg2017unified} analysis in Section~\ref{Feature analysis}. 

\subsection{Experimental Settings}
\label{Experimental Settings}
\subsubsection{Baselines}
We adopt four supervised learning models  \ac{SVM}, \ac{MLP}, \ac{GBDT}, and \ac{RNN} as our baseline.
The answer extraction task, which is designed as a binary classification problem to estimate the probability of a word being the answer, can be solved with binary classifier \ac{SVM},  \ac{MLP}, and \ac{GBDT} directly. 
For \ac{RNN}, owing to its iteration structure, we propose \ac{CRF} and cast the binary classification problem as a sequence tagging task to predict the label of each word in its sentence-level context.
This is to our knowledge the first time that \ac{CRF} has been applied to EEG-based tasks while prior works like relevance prediction~\cite{eugster2014predicting,eugster2016natural} with brain signals are often regarded as a binary classification problem.
In addition, we represent an untrained model where all its predictions are based on a random choice, which achieves the \ac{AUC} of 0.5. 
We do not feed other interactive features into our baseline since (1)~the strictly controlled user study does not involve mouse movements or eye movements to locate words. (2)~we expect our framework can adapt to situations such as internet surfing and off-line reading, where no content features such as queries submitted by the user are available.

On the other hand, for the answer sentence classification task, we consider \textit{perfectly relevant} sentences as positive samples.
Then, the answer sentence classification task is regarded as a classification problem of estimating the probability that a sentence being \textit{perfectly relevant}.
And it can also be treated as a ranking problem when ranking the corresponding sentences of a question accordingly. 
For the \ac{LR}, \ac{SVM}, and \ac{GBDT}, the probability that a sentence being positive is computed based on the predicted answer probability of each word in the answer extraction task. 
More specifically, the score $S$ of a sentence can be written as:
\begin{equation}
\label{eq:sentence_score}
S = \frac{max(W_1, ..., W_n) + mean(W_1, ..., W_n) + median(W_1, ..., W_n)}{3} 
\end{equation}
Where $W_i$ represents the score~(predicted probability) of the $i$-$th$ word in the given sentence,
$n$ refers to the number of words in this sentence.
According to Eq.~\ref{eq:sentence_score}, the score of a sentence is the average of the max/median/mean values of words' score in the sentence.
This method integrates word-level information to the sentence-level, which is similar to \citet{zheng2019human} in the attention estimation task with eye-tracking features. 
For the RNN model, \ac{EEG} features of each word in a sentence are fed into the network, and the final hidden layer is connected to a fully connected layer to obtain the probability distribution.
Then the sentence-level relevance labels are utilized to calculate the loss. 
Note that the \ac{CRF} module is not adopted in \ac{RNN} in the answer sentence classification task.
Finally, the untrained model, of which the procedure is the same as described in the answer extraction task, achieves the \ac{AUC} of 0.5 and \ac{MAP} of 0.615.
It is treated as the baseline without brain signals, and we calculate the margin of other models compared to it.  

\subsubsection{Data splitting strategies}
In our experiments, we perform two training strategies to deal with unseen questions and unseen users: \ac{CVOT} and \ac{LOPO}, respectively.  
The \ac{CVOT} strategy partitions the tasks and their corresponding sentences into ten folds, then uses the rest folds for training when validating each fold.
The \ac{LOPO} strategy learns a supervised model using the remaining participants' data when validating each participant.  

\subsubsection{Parameter setups}
We elaborate the parameters tuning procedures of UERCM and other baselines here.
For both models, the parameters are tuned according to the averaged \ac{AUC} in both data splitting strategies, i.e., user-independent and task-independent.

We train UERCM end-to-end with a min-batch size of 8 by using the Adam optimizer. 
For hyper parameters, the number of attention head, the hidden dimension, and the learning rate are selected from  $\{4,8\}$, $\{16,32\}$, and $\{10^{-4},10^{-3},10^{-2}\}$, respectively.
Besides, to accelerate the training procedure, we train UERCM on an NVIDIA TITAN XP 12G GPU and adopt the early-stop strategy when the validation performance does not improve after five iterations.
The implementation code of UERCM is based on PyTorch~\footnote{https://pytorch.org/}. 

For the baselines, the details are presented below:
(1)~For \ac{SVM}, we apply the radial basis function kernel and select the regularization parameter from $\{10^{-3},10^{-2},10^{-1},1,10^{1},10^{2},10^{3}\}$.
The kernel coefficient is automatically according to the data distribution
~\footnote{https://scikit-learn.org/stable/modules/generated/sklearn.svm.SVR.html}.
(2)~For \ac{MLP}, the learning rate, hidden dimension are selected from 
$\{10^{-4},10^{-3},10^{-2}\}$ and  $\{16,32,64\}$, respectively.
(3)~For \ac{GBDT}, the parameters include learning rate, estimator number, leaf
nodes, and the maximum tree depth, then the hyper parameters are selected from $\{10^{-4},10^{-3},10^{-2}\}$, $\{100,200,400\}$, $\{4,8\}$, and $\{4,8\}$, respectively.

\begin{figure}[h]
\vspace{-1mm}
  \centering
  \includegraphics[width=\linewidth]{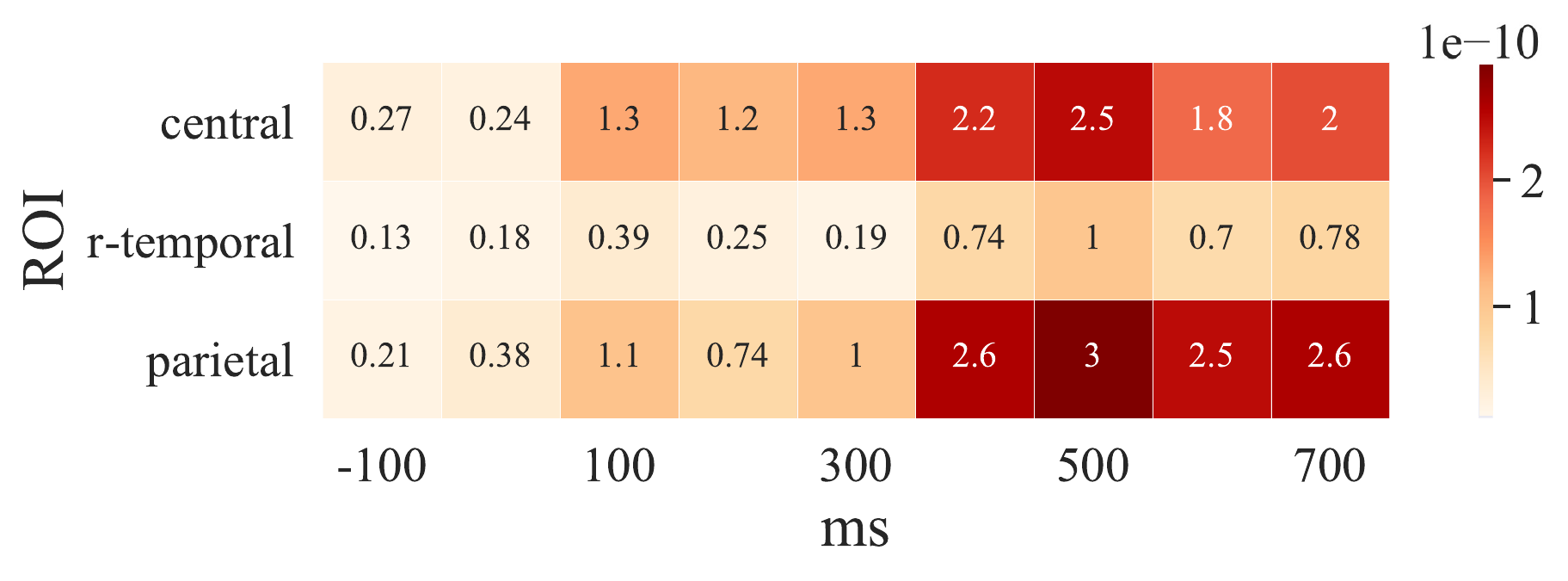}
  \caption{The average SHAP values. Higher SHAP value indicates higher feature importance.} 
  \label{fig:explainer}
  \Description[]{}
  \vspace{-3mm}
\end{figure}
 
\subsection{Feature analysis}
\label{Feature analysis}
To explore the effectiveness of \acp{ERPF} in different \ac{ROI} and time windows, we analyze the feature importance with \ac{SHAP}~\cite{lundberg2017unified}.
\ac{SHAP} is a unified method that can interpret feature importance in machine learning models.
Figure~\ref{fig:explainer} presents the average \ac{SHAP} values of the features with the \ac{LR} model in the answer extraction task.
Warmer color means higher \ac{SHAP} value, indicating the feature is more critical for answer words extraction. 
Results show that the contribution of feature importances in parietal and central areas is more than that in r-temporal.
Features with time points after 400ms have higher importance, which is not surprising since cognitive encoding of visual information takes time to happen, and N400 and P600 components in 320-750ms are supposed to be related to semantic understanding.
Besides, as an indicator of  ``matching process'' introduced in Section~\ref{statistical analysis:ERP Components}, N100 has higher feature importance than other components in adjacent time windows.